\documentclass[aps,reprint,pre,groupedaddress,longbibliography,amsmath]{revtex4-2}
\usepackage{graphicx,txfonts}
\usepackage[colorlinks,linkcolor=blue,citecolor=blue,anchorcolor=blue,urlcolor=blue]{hyperref}

\begin{document}

\title{Percolation transition of strongly connected clusters in finite dimensions and on complete graphs}

\date{\today}

\author{Qi Wang}
\affiliation{School of Physics, Hefei University of Technology, Hefei, Anhui 230009, China}

\author{Ming Li}
\email{lim@hfut.edu.cn}
\affiliation{School of Physics, Hefei University of Technology, Hefei, Anhui 230009, China}

\begin{abstract}
We study the percolation of strongly connected clusters (SCCs), in which sites are mutually reachable through directed paths, in systems with randomly oriented bonds by extensive simulations on hypercubic lattices from dimensions $d=2$ to $7$ and on complete graphs. Below the upper critical dimension $d_u=6$, critical SCCs exhibit a nontrivial fractal dimension $d_{\rm SCC}$ with a nonmonotonic dependence on dimension: $d_{\rm SCC}$ reaches a maximum at $d=4$, in contrast to the fractal dimension of ordinary percolation clusters, which increases monotonically with $d$. The SCC size distribution follows $\sim s^{-\tau_{\rm SCC}}$, where the Fisher exponent is consistent with the hyperscaling relation $\tau_{\rm SCC}=1+d/d_{\rm SCC}$ and, unlike $d_{\rm SCC}$ itself, increases monotonically with $d$. For $d\geq d_u$, mean-field behavior is recovered with $d_{\rm SCC}/d=1/3$, consistent with complete-graph results. However, unlike finite-dimensional lattices, complete graphs exhibit a double-scaling structure in the SCC size distribution: large SCCs are governed by the mean-field exponent $\tau_{\rm SCC}=4$, while small SCCs follow a distinct power law with exponent $\tau'=1$. At $d_u=6$, logarithmic corrections are also observed. These results show that SCCs remain well-defined fractal objects across dimensions, exhibiting both a nontrivial dimensional dependence and a nontrivial crossover from finite-dimensional to mean-field behavior.
\end{abstract}

\maketitle

\section{Introduction}

Geometrical structure is one of the important perspectives in establishing a unified understanding of critical phenomena~\cite{Ma2018}. Near criticality of a continuous phase transition, fluctuations occur on all length scales and give rise to self-similar spatial patterns characterized by universal critical exponents and fractal geometry. Percolation, which considers the connectivity of sites on a lattice or graph when bonds are occupied randomly with probability $p$, provides one of the clearest realizations of such geometrical criticality, where the emergence of a spanning cluster is accompanied by a divergent correlation length and nontrivial fractal organization~\cite{Stauffer1994}. Through the Fortuin-Kasteleyn (FK) representation~\cite{Kasteleyn1969,Fortuin1972}, the partition function of the Potts model can be reformulated in terms of random clusters, thereby extending geometrical interpretations of criticality to a broad class of thermodynamic phase transitions. Understanding how connectivity structures evolve at criticality has therefore long been a central theme in statistical physics.

The geometry of critical clusters is commonly characterized by their fractal dimensions. For example, the largest FK cluster scales as $C_1\sim L^{d_f}$, where $d_f$ is its fractal dimension. Beyond cluster size, the nature of a phase transition is often encoded in finer connectivity structures, including hulls, external perimeters, shortest paths, and backbones~\cite{Stauffer1994}. These geometrical substructures possess their own universal scaling properties. In two dimensions (2D), remarkable progress has been achieved through Coulomb gas theory~\cite{Nienhuis1984,Nienhuis1987}, conformal invariance~\cite{DiFrancesco1997}, and Schramm--Loewner evolution~\cite{Kagery2004,Cardy2005}, which established exact relations between critical interfaces and their fractal geometry.

For example, the fractal dimension, hull exponent, and external perimeter exponent of 2D percolation take the exact values $d_f=91/48$, $d_{\mathrm{hull}}=7/4$, and $d_{\mathrm{EP}}=4/3$, respectively~\cite{Coniglio1989,Smirnov2001}. Recently, the backbone exponent $d_{\mathrm{B}}$, which characterizes the fractal geometry of biconnected clusters, has also been exactly solved using the conformal loop ensemble~\cite{Nolin2023,Nolin2025}. Different from other exactly solvable critical exponents, the backbone exponent is a transcendental number rather than a rational number. For percolation, it gives $d_{\mathrm{B}}=1.6433331\ldots$, which is consistent with the previously reported Monte Carlo estimate $d_{\rm B}=1.64339(5)$~\cite{Fang2022}. Above dimension $d=2$, exact solutions are available only for the mean-field behaviors in infinite dimensions or above $d=10$ dimensions~\cite{Hara1990,Grimmett1999,Fitzner2017}, and thus the crossover from low-dimensional criticality to mean-field behavior remains largely understood through finite-size scaling analyses.

Another important refinement of connectivity arises from directionality. When bonds are endowed with preferred directions, connectivity becomes path dependent and generally anisotropic. Studies of directed percolation have shown that directional constraints can fundamentally modify critical behavior, leading to new universality classes characterized by distinct longitudinal and transverse correlation-length exponents~\cite{Domany1981,Essam1988,Grassberger1989,Wang2013}. As the directional bias is reduced, the system may cross over toward isotropic percolation. Simulation results indicate that anisotropic fixed points are stable against generic directional perturbations, whereas isotropic percolation is recovered only in the fully symmetric limit~\cite{Zhou2012}. These results demonstrate that introducing directionality enriches percolation criticality beyond purely geometric connectivity, which makes percolation closely related to many other issues, such as turbulence~\cite{Hof2022}, glass~\cite{Shrivastav2016}, the Leidenfrost effect~\cite{Chantelot2021}, quantum circuits~\cite{Piroli2023}, and Janus systems~\cite{Hu2022}.

\begin{figure}
\centering
\includegraphics[width=\columnwidth]{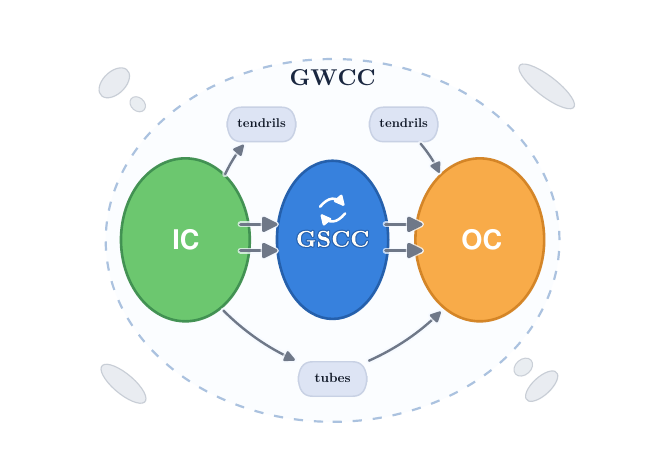}
\caption{Sketch of the bow-tie structure of a directed graph. The giant SCC (labeled as GSCC in the sketch) consists of sites that are mutually reachable through directed paths. The IC contains sites that can reach an SCC but cannot be reached from it, while the OC contains sites that are reachable from an SCC but do not necessarily have return access to it. The giant SCC together with its associated IC is often referred to as the giant IC, while the giant SCC together with its associated OC is referred to as the giant OC. Around these giant clusters, there are also small connecting structures such as tubes and tendrils. Omitting bond directions, these structures form the giant WCC. In addition, some finite clusters are also observed, which are disconnected from these giant clusters.}    \label{f1}
\end{figure}

Beyond one-way reachability, directed systems also support the stronger notion of mutual accessibility (Fig.~\ref{f1}). The subset of sites that can mutually reach one another through directed paths forms a strongly connected cluster (SCC)~\cite{Tarjan1972,Broder2000}. Sites that can reach an SCC through directed paths, but cannot be reached from it, form the associated in-cluster (IC). Conversely, sites that are reachable from the SCC but do not have return access to it belong to the associated out-cluster (OC). Treating all directed bonds as undirected gives the usual weakly connected cluster (WCC). When an extensive SCC emerges, it constitutes the giant SCC. The giant SCC together with its associated IC is often referred to as the giant IC, while the giant SCC together with its associated OC is referred to as the giant OC. Thus, the giant SCC is given by the intersection of the giant IC and OC. Around these giant structures, as illustrated in Fig.~\ref{f1}, tubes and tendrils may also appear.

Similar to backbones, the SCC also represents a form of redundant connectivity through mutual reachability. However, its critical geometry is distinct from that of the backbone. Numerical studies in 2D have shown that the fractal dimension of SCCs, $d_{\mathrm{SCC}}\approx1.80$~\cite{Noronha2018}, is larger than that of the backbone $d_{\mathrm{B}}\approx1.643$~\cite{Nolin2025}, yet remains smaller than that of the full critical percolation cluster $d_f=91/48\approx1.895$. A recent numerical study further suggests that 2D SCCs have the same hull fractal dimension, $d_{\rm hull}=4/3$, as backbones~\cite{Newman2026}. In network-based and mean-field settings, directed systems have attracted even broader attention, as they refer to many network systems~\cite{Dorogovtsev2001,Schwartz2002,Boguna2005,AngelesSerrano2007,Kryven2016,vanIeperen2023,Li2021,Costa2025,Garofalo2026}. On complete graphs, the percolation of SCCs can be solved rigorously~\cite{Luczak1990,Luczak2009}, providing an exact description of the corresponding complete-graph mean-field behavior. Because complete graphs do not have a linear system size, their fractal properties are characterized by the volume scaling, $\sim V^{d_{\mathrm{SCC}}^*}$, with $d_{\mathrm{SCC}}^*=1/3$, which coincides with that of backbones~\cite{Huang2018}. In contrast, standard percolation has $d_f^*=2/3$ in the same volume-based notation.

Despite these advances, a unified understanding of SCC criticality across dimensions is still lacking. In particular, how the fractal geometry of SCCs evolves from low-dimensional lattices to the mean-field limit remains unknown. It is also unclear whether IC and OC display analogous critical scaling and how their behavior is related to that of the SCC itself. These questions motivate the present work.

In this paper, we investigate directed bond percolation on hypercubic lattices in dimensions $d=2$–$7$ and on complete graphs. Bond orientations are introduced through a symmetric construction such that the percolation threshold of SCCs coincides with that of standard bond percolation in all dimensions.

For $d < 6$, our numerical results show that the SCC fractal dimension lies in the range of $d_{\rm B} < d_{\rm SCC} < d_f$. The dimensional dependence of $d_{\rm SCC}$ is nonmonotonic, with a maximum at $d=4$. This could be a result of the competition between enhanced connectivity with increasing dimension and the increasingly tree-like structure of critical clusters, which suppresses the loop structures required for mutual accessibility in SCCs. At $d=6$, it reduces to the mean-field value $d_{\rm SCC} =2$, with multiplicative logarithmic corrections, suggesting the percolation transition of SCCs shares the same upper critical dimension $d_u=6$ as standard percolation.

Above $d_u=6$, it takes the complete-graph mean-field value $d_{\rm SCC} = d_{\rm B} = d/3$. However, the SCC behavior on high-dimensional lattices remains distinct from that on complete graphs. In particular, the total number of SCCs scales proportionally to the system volume $V=L^d$ in finite-dimensional lattices, whereas it grows only as $\sim \ln V$ on complete graphs, leading to a qualitatively different critical cluster-size distribution. For $d=7$, the size distribution of SCCs takes the Fisher exponent predicted by the hyperscaling value $\tau_{\rm SCC}= 1 +d/d_{\rm SCC}=4$. For complete graphs, the mean-field value $\tau_{\rm SCC}=4$ governs only large SCCs, while small SCCs follow a different scaling characterized by an effective exponent $\tau'=1$. We further show that, at the SCC threshold, both the giant IC and OC are fractal and possess the same fractal dimension as standard percolation.

The remainder of the paper is organized as follows. In Sec.~\ref{sec:model}, we define the model, observables, and numerical methods. In Sec.~\ref{sec:results}, we present results for hypercubic lattices in dimensions $d=2$--$7$, together with the complete-graph limit. Finally, Sec.~\ref{sec:discussion} summarizes our findings and discusses future perspectives.

\section{Models, observables, and algorithm}   \label{sec:model}

\begin{figure}
\centering
\includegraphics[width=\columnwidth]{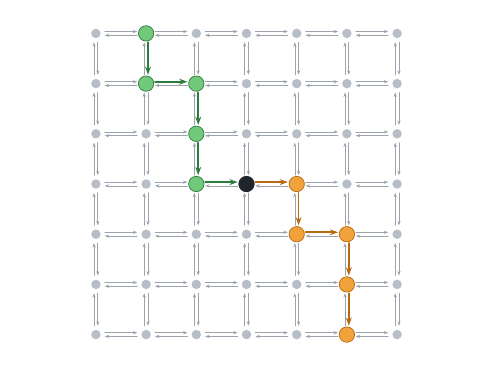}
\caption{Illustration of SCC percolation on a directed square lattice. Each nearest-neighbor pair of sites has two antiparallel directed bonds. Each directed bond is independently occupied with probability $p$. Starting from a reference site (the black site at the center), the yellow path illustrates an outgoing path and the green path an incoming path. For clarity, other occupied bonds in the configuration are omitted. If the reference site belongs to a spanning SCC, both paths must extend to macroscopic distances. Since the two classes of bonds are independently occupied, the spanning of either class is equivalent to a standard bond-percolation process on the lattice. Therefore, the percolation threshold of SCCs coincides with that of standard percolation.}    \label{f2}
\end{figure}

\subsection{Directed bond percolation}

We consider $d$-dimensional hypercubic lattices of linear size $L$ with periodic boundary conditions. Each nearest-neighbor pair of sites is connected by two antiparallel directed bonds, as sketched in Fig.~\ref{f2}. The total number of directed bonds is $E=2dV$, with $V=L^d$ the system volume, which is twice that of the corresponding undirected lattice. Percolation configurations are generated by occupying each directed bond independently with probability $p$. This construction naturally extends to complete graphs. In this case, each pair of distinct sites is connected by two antiparallel directed bonds.

With this setting, the percolation threshold of SCCs coincides with the critical point $p_c$ of standard bond percolation on the corresponding undirected lattice/graphs (simulation results and detailed arguments will be given in Sec.~\ref{sec-pt}). In our simulations, we use the known critical thresholds $p_c=1/2$ for $d=2$~\cite{Stauffer1994}, $p_c=0.24881182$ for $d=3$~\cite{Wang2013}, and $p_c=0.16013122,\,0.11817145,\,0.0942019,\,0.0786752$ for $d=4$--$7$, respectively~\cite{Mertens2018}. For complete graphs, the critical point is given by $p_c=1/(V-1)$, corresponding to a critical average in-degree or out-degree $q_c=p_c(V-1)=1$~\cite{Luczak1990,Dorogovtsev2001,Luczak2009}. Such a critical configuration is a directed Erd\H{o}s-R\'enyi random graph~\cite{Bollobas2001} with average in-degree or out-degree $q_c=1$.

\subsection{Observables}

For directed-bond configurations at $p_c$, we measure the following quantities:
\begin{itemize}
  \item Let $\mathcal{C}_1(\mathrm{X})$ denote the size of the largest connected cluster of type $\mathrm{X}$ in a given realization, where $\mathrm{X}\in\{\mathrm{SCC}, \mathrm{IC}, \mathrm{OC}\}$. The ensemble-averaged size is defined as
      \begin{equation}
      C_1(\mathrm{X}) \equiv \langle \mathcal{C}_1(\mathrm{X}) \rangle.
      \end{equation}

  \item The probability density $P(x)$ of the rescaled largest SCC size,
      \begin{equation}
      x \equiv \frac{C_1}{L^{d_{\mathrm{SCC}}}},
      \end{equation}
        where $d_{\mathrm{SCC}}$ is the fractal dimension of SCCs. For complete graphs, the finite-size scaling is respect to volume $V$, so that
      \begin{equation}
      x \equiv \frac{C_1}{V^{d_{\mathrm{SCC}}^*}},
      \end{equation}
        where $d_{\mathrm{SCC}}^*$ is the volume fractal dimension. For $d$-dimensional hypercubic lattices, it has
      \begin{equation}
      d_{\mathrm{SCC}}^* = \frac{d_{\mathrm{SCC}}}{d}.
      \end{equation}

  \item The susceptibility-like quantity,
      \begin{equation}
      \chi\equiv\frac{1}{V}\left\langle\sum_{i\ne1}\mathcal{C}_i^2\right\rangle,
      \label{eq_order_chi}
      \end{equation}
      where the sum runs over all cluster sizes $\mathcal{C}_i$ with the largest one excluded.

  \item The mean number of SCCs,
      \begin{equation}
      N_{\mathrm{SCC}} \equiv \langle \mathcal{N}_{\mathrm{SCC}} \rangle,
      \end{equation}
      where $\mathcal{N}_{\mathrm{SCC}}$ is the total number of SCCs in a given configuration.

  \item The cluster-size distribution of SCCs,
      \begin{equation}
      n(s,L) = \frac{\langle \mathcal{N}_s(\Delta s) \rangle}{V \, \Delta s},
      \end{equation}
      where $\mathcal{N}_s(\Delta s)$ is the number of SCCs with sizes in the interval $[s, s+\Delta s)$. We use geometrical binning with $\Delta s = a^k$ for the $k$-th bin starting from $s=1$, where $a>1$ controls the binning resolution. The bin center is defined as $s'=\sqrt{s(s+\Delta s)}$.
  \item The Binder cumulant is calculated as
      \begin{equation}
      Q=\frac{\langle\mathcal{S}_2\rangle^2}{3\langle\mathcal{S}_2^2\rangle-2\langle\mathcal{S}_4\rangle},
     \end{equation}
     where $\mathcal{S}_2=\sum_i\mathcal{C}_i^2$ is the quadratic sum of SCCs in a configuration, and $\mathcal{S}_4=\sum_i\mathcal{C}_i^4$ is the fourth-power sum of SCCs.
\end{itemize}
Here, $\langle \cdot \rangle$ denotes averaging over independent realizations.

\subsection{Tarjan's algorithm}

Identifying SCCs in a directed graph is a classical problem in graph theory. Several linear-time algorithms are available, including Tarjan's algorithm~\cite{Tarjan1972}, the Kosaraju--Sharir algorithm~\cite{Sharir1981,Sedgewick2001}, and Gabow's algorithm~\cite{Cheriyan1996,Gabow2000}.

In this work, we employ Tarjan's algorithm, which performs a single depth-first search over the system. Each site $v$ is assigned a discovery index $dfn[v]$ and a low-link value $low[v]$. The discovery index $dfn[v]$ records the order in which sites are visited, while the low-link value $low[v]$ denotes the smallest discovery index reachable from $v$ via the depth-first-search subtree and at most one back edge. When the condition $dfn[v] = low[v]$ is satisfied, $v$ is the root of an SCC. After finding all the roots of SCCs, which runs in $O(V+E)$ time and requires $O(V)$ auxiliary memory, we can identify all the SCCs. Then, we construct a compressed sparse row representation of the directed graph~\cite{Broder2000}. Starting from the largest SCC (i.e., the giant SCC), forward and reverse traversals on this structure allow us to efficiently identify the giant IC and OC.

\subsection{Fit protocol}

Our main focus is the fractal dimension of the SCC, IC, and OC at criticality. To estimate the fractal dimension, we fit the mean size $C_1(\mathrm{X})$ using the finite-size scaling ansatz
\begin{equation}
C_1(\mathrm{X})=L^{d_\mathrm{X}}\left(a_0+a_1L^{-\omega_1}+\cdots\right),
\label{eq_fit}
\end{equation}
where $d_\mathrm{X}$ is the fractal dimension associated with cluster type $\mathrm{X}$, and $\omega_1$ is the leading correction-to-scaling exponent.

In the fitting procedure, a lower cutoff $L\geq L_{\min}$ is imposed. We assess the quality and stability of the fits by systematically increasing $L_{\min}$. Our preferred estimates correspond to the smallest $L_{\min}$ for which the Chi-square per degree of freedom is of order unity and the fitted parameters remain stable upon further increasing $L_{\min}$.

In practice, allowing $\omega_1$ to vary freely often leads to unstable fits. We therefore perform fits with $\omega_1$ fixed to several plausible values and estimate the final results by comparing the corresponding fits. The spread of these results is taken as an additional contribution to the systematic uncertainty.

For complete graphs, there is no intrinsic linear size $L$. Finite-size scaling is therefore formulated in terms of the system volume $V$. Accordingly, the scaling ansatz becomes
\begin{equation}
C_1(\mathrm{X})=V^{d_\mathrm{X}^*}\left(a_0+a_1V^{-\omega_1}+\cdots\right).
\end{equation}

\section{Results}   \label{sec:results}

In this section, we first investigate the percolation of SCCs on hypercubic lattices in dimensions $d=2$-$7$, focusing on their fractal dimensions and their dependence on spatial dimension. We then examine the multiplicative logarithmic corrections at the upper critical dimension $d_u=6$, followed by a comparison between the high-dimensional lattice results and those on complete graphs, with particular emphasis on their distinct finite-size scaling behaviors.

\subsection{Hypercubic lattices}

\subsubsection{Percolation threshold}  \label{sec-pt}

\begin{figure}
\centering
\includegraphics[width=\columnwidth]{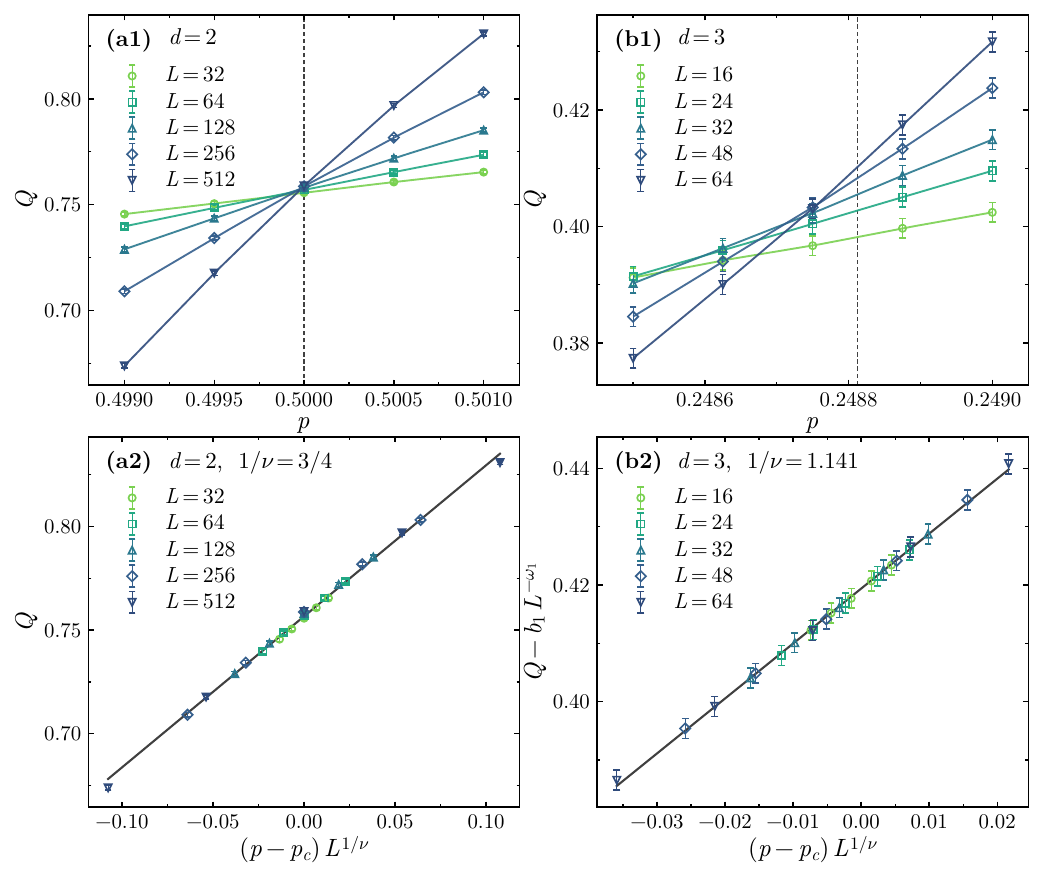}
\caption{Finite-size scaling of the Binder cumulant $Q$ for (a) $d=2$ and (b) $d=3$. Panels (a1,b1) show the Binder cumulant $Q$ as a function of $p$ near criticality for different system sizes $L$. The dashed lines indicate the percolation thresholds $p_c=1/2$ (2D) and $p_c=0.24881182$ (3D). Panels (a2,b2) present the corresponding data collapse over a broader range of $p$ by rescaling the horizontal axis as $(p-p_c)L^{1/\nu}$, where $\nu$ is the correlation-length exponent of standard percolation. For panel (b2), finite-size corrections are taken into account by shifting the Binder cumulant as $Q-b_1L^{-\omega_1}$, with $\omega_1\approx0.6$ and $b_1\approx-0.11$, resulting in an improved data collapse.}  \label{f3}
\end{figure}

Near criticality, the Binder cumulant $Q$ follows the finite-size scaling form~\cite{Qian2005,Deng2005},
\begin{equation}
Q(p,L) = Q_c + \sum_{k=1} a_k (p-p_c)^k L^{k/\nu} + \sum_{k=1} b_k L^{-\omega_k},   \label{eq-qpl}
\end{equation}
where $Q_c$ is the universal critical value in the thermodynamic limit. The terms $(p-p_c)^k L^{k/\nu}$ describe the scaling behavior approaching criticality, while $L^{-\omega_k}$ accounts for finite-size corrections at $p_c$. Equation~\eqref{eq-qpl} implies that Binder cumulants for different system sizes tend to intersect at the critical point as $L\to\infty$.

In Fig.~\ref{f3}, we show the cases $d=2$ and $3$ as representative examples. For $d=2$, shown in Fig.~\ref{f3}(a1), the curves of $Q(p,L)$ for different system sizes intersect at $p_c=1/2$, indicating that SCC percolation shares the same threshold as standard bond percolation. Furthermore, plotting $Q(p,L)$ against $(p-p_c)L^{1/\nu}$ with the 2D exponent $\nu=4/3$ yields an excellent data collapse [Fig.~\ref{f3}(a2)], suggesting that SCC percolation also shares the same correlation-length exponent $\nu$.

For $d=3$, shown in Fig.~\ref{f3}(b1), the crossings of $Q(p,L)$ exhibit noticeable finite-size drift, but clearly converge toward the standard percolation threshold. To obtain a satisfactory data collapse, finite-size corrections must therefore be included through the term $L^{-\omega_k}$ in Eq.~\eqref{eq-qpl}. Fitting the data to Eq.~\eqref{eq-qpl} with $a_k=0$ and $b_k=0$ for $k\geq2$, we find $\omega_1=0.6(2)$. Using this value, a good data collapse is obtained by plotting $Q(p,L)-b_1L^{-\omega_1}$ against $(p-p_c)L^{1/\nu}$ with the 3D percolation exponent $1/\nu\approx 1.141$~\cite{Wang2013a,Xu2013}, see Fig.~\ref{f3}(b2). It is pointed out that the value $\omega_1=0.6(2)$ is also consistent with that of standard percolation in 3D~\cite{Deng2005}.

The same behavior is observed in other dimensions, strongly suggesting that SCC percolation in our model shares both the percolation threshold and the correlation-length exponent $\nu$ of standard percolation. To understand this behavior, we introduce a two-arm picture by dividing all directed bonds into two complementary classes, referred to as out- and in-bonds. As illustrated in Fig.~\ref{f2}, starting from a reference site, a bond pointing away from the site is defined as an out-bond, while the opposite bond is defined as an in-bond. These labels are purely relative to the reference site and should not be directly identified with ICs or OCs.

\begin{figure}
\centering
\includegraphics[width=\columnwidth]{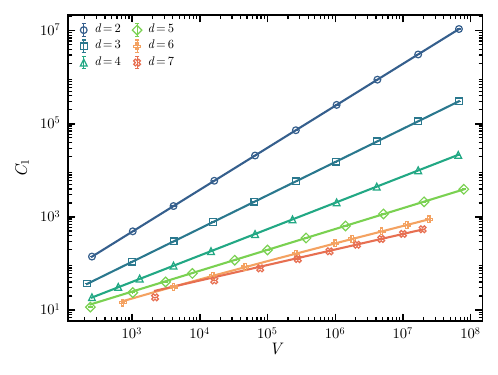}
\caption{Finite-size scaling of the largest SCC on hypercubic lattices. The mean size $C_1$ is plotted versus the system volume $V=L^d$ for $d=2$--$7$ at $p=p_c$. Solid lines indicate the fitted scaling forms, $C_1\sim V^{d_{\mathrm{SCC}}^*}$, with $d_{\mathrm{SCC}}^*=d_{\mathrm{SCC}}/d$ listed in Table~\ref{taball}.} \label{f4}
\end{figure}

If the SCC spans the system ($p \geq p_c$), directed paths starting from the reference site must both reach and return from macroscopic distances. This implies the simultaneous existence of at least two paths extending to macroscopic distances, formed exclusively by out- and in-bonds, respectively (see Fig.~\ref{f2}). Equivalently, the two classes of bonds must each span the lattice independently. Since out- and in-bonds are independently occupied, the spanning of either class is equivalent to a standard bond-percolation process on the lattice. Therefore, the percolation threshold of SCCs coincides with that of standard percolation. This argument applies to arbitrary lattices and graphs. Similar arguments have also been used to determine SCC thresholds in related models~\cite{Redner1982,Zhou2012,Noronha2018}, although the resulting thresholds do not always coincide with those of standard percolation because of different model definitions.

In addition, if all directed bonds are treated as undirected, the effective connectivity probability of two neighboring sites becomes $p_{\mathrm{eff}} = 1-(1-p)^2$. At $p=p_c$, one has $p_{\mathrm{eff}} > p_c$, implying that the WCC is already in the supercritical phase. Therefore, no critical behavior is associated with WCCs when SCCs percolate.

\subsubsection{Fractal dimensions of SCCs}

Figure~\ref{f4} shows the size of the largest SCC, $C_1$, at criticality as a function of the system volume $V=L^d$ for dimensions $d=2$-$7$. In all finite dimensions, $C_1$ exhibits a clear power-law scaling,
\begin{equation}
C_1 \sim L^{d_{\mathrm{SCC}}} \sim V^{d_{\mathrm{SCC}}^*},
\end{equation}
where $d_{\mathrm{SCC}}^*=d_{\mathrm{SCC}}/d$. This indicates a continuous percolation transition of SCCs, with fractal dimension $d_{\mathrm{SCC}}$.

To extract $d_{\mathrm{SCC}}$, we fit the data to the finite-size scaling ansatz in Eq.~\eqref{eq_fit}, excluding $d=6$ where multiplicative logarithmic corrections are present (to be discussed in Sec.~\ref{sec-d6}). The resulting estimates are summarized in Table~\ref{taball}. As a consistency check, we also analyze the scaling of the second-largest SCC, which yields compatible results. The corresponding scaling forms $\sim V^{d_{\mathrm{SCC}}/d}$ are shown as solid lines in Fig.~\ref{f4}, and are in good agreement with the numerical data for increasing $V$.

\begin{table}
\caption{Estimated fractal dimensions of SCCs, ICs, and OCs on hypercubic lattices and complete graphs. For $d<6$, the results are obtained from fits to the finite-size scaling ansatz in Eq.~\eqref{eq_fit}, while at the upper critical dimension $d_u=6$ multiplicative logarithmic corrections are taken into account using Eq.~\eqref{eq_fitln}. For complete graphs (CG), finite-size scaling is formulated in terms of $V$, and the results are the volume fractal dimension $d_{\mathrm{X}}^*$. For comparison, we also list the known results of $d_f$ and $d_\mathrm{B}$. For $d \ge d_u=6$, the complete-graph mean-field results are $d_f = 2d/3$~\cite{Kenna2017,Huang2018,Li2024,Lu2024}, and $d_{\mathrm{B}}=d/3$~\cite{Luczak1990,Luczak2009,Huang2018}.}     \label{taball}
\begin{ruledtabular}
\begin{tabular}{cllll}
$d$   & \multicolumn{1}{c}{$d_{\mathrm{SCC}}$}  & \multicolumn{1}{c}{$d_{\mathrm{IC}}=d_{\mathrm{OC}}$}  & \multicolumn{1}{c}{$d_f$}   & \multicolumn{1}{c}{$d_{\mathrm{B}}$}  \\
\hline
$2$   &  $1.8041(3)$   &  $1.895(2)$   &  $91/48$~\cite{Coniglio1989,Smirnov2001}   &  $1.6433\ldots$~\cite{Nolin2023,Nolin2025}                     \\
$3$   &  $2.1432(7)$   &  $2.525(7)$   &  $2.52293(10)$~\cite{Xu2013}     &  $1.855(15)$~\cite{Rintoul1994}       \\
$4$   &  $2.266(5)$    &  $3.042(4)$   &  $3.0446(7)$~\cite{Zhang2021}    &  $1.9844(11)$~\cite{Zhang2021}        \\
$5$   &  $2.241(5)$    &  $3.52(3)$    &  $3.5260(14)$~\cite{Zhang2021}   &  $2.0226(27)$~\cite{Zhang2021}        \\
$6$   &  $1.998(8)$    &  $4.02(3)$    &  $4$                             &  $2$                                    \\
$7$   &  $2.332(4)$    &  $4.66(1)$    &  $14/3$                          &  $7/3$                                  \\
\hline
CG    & $0.3336(4)$ & $0.667(2)$ & $0.6664(5)$~\cite{Huang2018} & $0.3337(15)$~\cite{Huang2018}      \\
Theory     & $1/3$       &  $2/3$         &  $2/3$  &  $1/3$
\end{tabular}
\end{ruledtabular}
\end{table}

\begin{figure}
\centering
\includegraphics[width=\columnwidth]{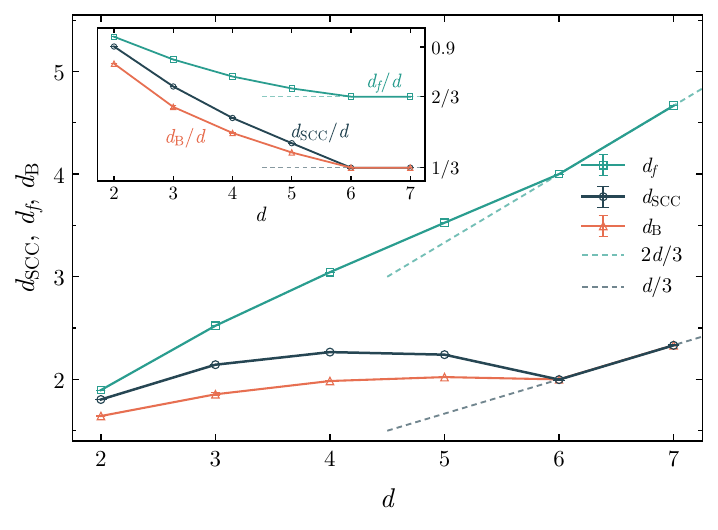}
\caption{The fractal dimension $d_{\mathrm{SCC}}$ as a function of spatial dimension $d$, compared with standard percolation cluster $d_f$ and backbone $d_\mathrm{B}$. The values of $d_{\mathrm{SCC}}$, $d_f$ and $d_\mathrm{B}$ are as listed in Table~\ref{taball}. The dashed lines represent the mean-field values $2d/3$ and $d/3$. The inset shows that $d_{\mathrm{SCC}}/d$, $d_f/d$, and $d_\mathrm{B}/d$ decrease monotonically with increasing $d$ and approach their mean-field limits, $2/3$ or $1/3$, for $d \ge d_u=6$.}   \label{f5}
\end{figure}

\begin{figure}[t]
\centering
\includegraphics[width=\columnwidth]{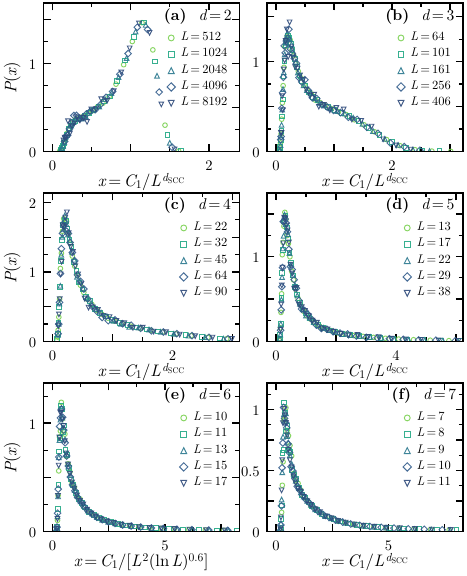}
\caption{Scaling collapse of the probability density $P(x)$ for the rescaled size of the largest SCC $x$. For $d\neq6$, $x\equiv C_1/L^{d_{\mathrm{SCC}}}$, with the fitted values of $d_{\mathrm{SCC}}$ listed in Table~\ref{taball}. At $d=6$, a multiplicative logarithmic correction is present, and $x\equiv C_1/[L^{d_{\mathrm{SCC}}}(\ln L)^w]$ as discussed in Sec.~\ref{sec-d6}.} \label{f6}
\end{figure}

We find that $d_{\mathrm{SCC}}$ exhibits a nonmonotonic dependence on the spatial dimension $d$. As shown in Fig.~\ref{f5}, $d_{\mathrm{SCC}}$ first increases with $d$, reaches a maximum at $d=4$, and then decreases as $d$ increases further. Upon approaching $d=6$, it crosses over to the complete-graph mean-field behavior $d_{\mathrm{SCC}}=d/3$~\footnote{The finite-size scaling above the upper critical dimension is subtle and depends on the boundary conditions~\cite{Kenna2017,Li2024,Lu2024}. Under periodic boundary conditions, the fractal dimensions $d_f=2d/3$ and $d_{\mathrm{SCC}}=d_{\mathrm{B}}=d/3$ correspond to the so-called complete-graph mean-field behavior. In contrast, dimension-independent values such as $d_{\mathrm{SCC}}=d_{\mathrm{B}}=2$ and $d_f=4$ are associated with Gaussian mean-field behavior, which can be observed under free boundary conditions. We use periodic boundary conditions throughout this work, for which the complete-graph mean-field scaling is the relevant behavior.}, beyond which it becomes an increasing function of $d$ again. This behavior provides further evidence that the upper critical dimension of SCC percolation is $d_u=6$, consistent with standard percolation. This is in contrast to directed percolation, whose upper critical dimension is $d_u=4$ due to its intrinsic anisotropy~\cite{Wang2013}, whereas SCC percolation remains isotropic.

This nonmonotonic dependence on the spatial dimension $d$ can be qualitatively understood as resulting from two competing mechanisms in the formation of SCCs. First, as $d$ increases, the number of available directions and alternative connecting paths increases, which generally facilitates connectivity and tends to make SCCs more compact, thereby increasing $d_{\mathrm{SCC}}$. On the other hand, unlike standard percolation, SCCs require mutual accessibility and therefore rely on loop structures that allow paths to return to their starting regions. Critical clusters become increasingly tree-like with increasing $d$, which tends to suppress such mutual connectivity and reduce $d_{\mathrm{SCC}}$. The competition between these two mechanisms may therefore give rise to the observed nonmonotonic dependence of $d_{\mathrm{SCC}}$ on $d$.

We emphasize, however, that this argument does not explain why the maximum of $d_{\mathrm{SCC}}$ occurs specifically at $d=4$, nor does it establish that a maximum must occur. Similar competing mechanisms may also operate in other geometrical structures, such as backbone clusters. Indeed, the existing numerical results summarized in Table~\ref{taball} suggest that the fractal dimension of backbones reaches a maximum around $d=5$. Thus, nonmonotonic dimensional dependence appears to be a broader feature of geometrical connectivity, while the different locations of the maxima for SCCs and backbones remain to be understood. Nevertheless, for $d<6$, we consistently find $d_\mathrm{B} < d_{\mathrm{SCC}}<d_f$. This hierarchy reflects the progressively stronger connectivity constraints imposed on the clusters, from ordinary connectivity to mutual connectivity and backbone connectivity. The rescaled quantity $d_{\mathrm{SCC}}^*=d_{\mathrm{SCC}}/d$, however, decreases monotonically with increasing $d$, exhibiting the same overall trend as $d_\mathrm{B}$ and $d_f$ (see the inset of Fig.~\ref{f5}).

To validate the extracted values of $d_{\mathrm{SCC}}$, we further examine the probability density $P(x)$ of the rescaled variable $x \equiv C_1/L^{d_{\mathrm{SCC}}}$. As shown in Fig.~\ref{f6} for $d=2$-$5$, and $7$, with $d_{\mathrm{SCC}}$ in Table~\ref{taball}, data for different system sizes collapse onto a single curve, further supporting the scaling ansatz.

\begin{figure}
\centering
\includegraphics[width=\columnwidth]{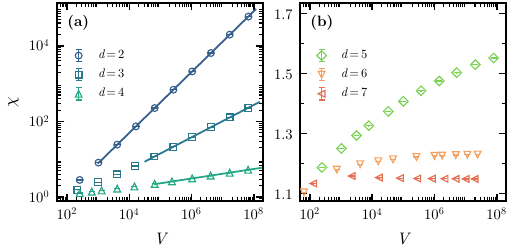}
\caption{Finite-size behavior of the susceptibility $\chi$ at $p=p_c$ as a function of $V=L^d$. (a) For $d=2$-$4$, $\chi$ increases algebraically with $V$; solid lines show the scaling of $\chi\sim L^{2d_{\mathrm{SCC}}-d} \sim V^{2d_{\mathrm{SCC}}^*-1}$. (b) For $d=5$-$7$, $\chi$ approaches a finite value as $V$ increases.} \label{f7}
\end{figure}

The finite-size scaling theory gives the susceptibility-like quantity $\chi$ the scaling form
\begin{equation}
\chi \sim L^{2d_{\mathrm{SCC}}-d} \sim V^{2d^*_{\mathrm{SCC}}-1}.   \label{eq-chi}
\end{equation}
Using the values of $d_{\mathrm{SCC}}$ in Table~\ref{taball}, the exponent $2d_{\mathrm{SCC}}-d$ is positive for $d\leq4$, indicating that $\chi$ diverges as $L\to\infty$, as expected for a continuous phase transition. In contrast, $2d_{\mathrm{SCC}}-d<0$ for $d>4$, implying that $\chi$ remains finite at criticality. In Fig.~\ref{f7}, we show $\chi$ for $d=2$-$7$. For $d=2$--$4$, the data are consistent with the scaling Eq.~(\ref{eq-chi}), whereas for $d>4$, $\chi$ approaches a finite value as $L$ increases, in agreement with the prediction. We point out that a finite $\chi$ at criticality does not imply a breakdown of the continuous transition, similar behaviors are also observed in percolation on scale-free networks~\cite{Zhao2025}.

\subsubsection{The size distribution of SCCs}

\begin{figure*}
\centering
\includegraphics[width=\textwidth]{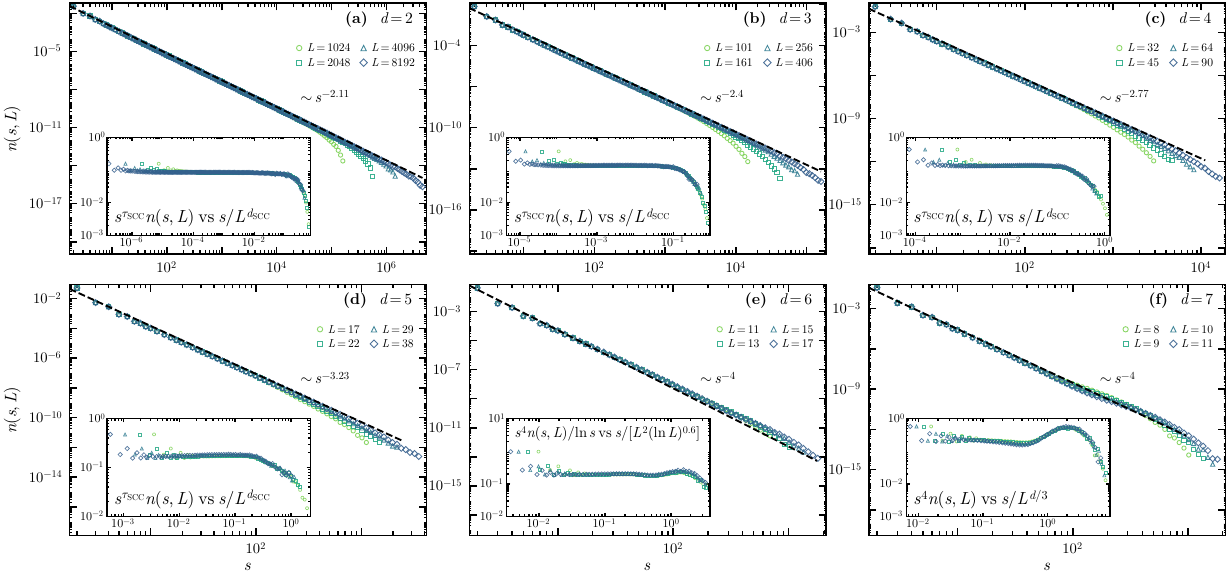}
\caption{Cluster-size distributions $n(s,L)$ of finite SCCs at criticality for $d=2$-$7$. Dashed lines show the scaling $\sim s^{-\tau_{\mathrm{SCC}}}$, where $\tau_{\mathrm{SCC}}=1+d/d_{\mathrm{SCC}}$ for $d=2$-$5$ and $\tau_{\mathrm{SCC}}=4$ for $d=6,7$. Insets show $s^{\tau_{\mathrm{SCC}}}n(s,L)$ versus $s/L^{d_{\mathrm{SCC}}}$ for $d\neq6$, and $s^4n(s,L)/\ln s$ versus $s/[L^2(\ln L)^{0.6}]$ for $d=6$.}
\label{f8}
\end{figure*}

Besides the largest SCC, we now examine the size distribution of SCCs at criticality. In standard percolation, the cluster-size distribution follows a power-law form
\begin{equation}
n(s,L) = s^{-\tau} \tilde{n}(s/L^{d_f}).
\end{equation}
Here, the scaling function $\tilde{n}(x)$ is almost a constant for $s\ll L^{d_f}$, and vanishes for $s\gg L^{d_f}$. A similar scaling behavior is expected for SCCs.

In Fig.~\ref{f8}, we plot the size distribution $n(s,L)$ of SCCs for $d=2$-$7$ at $p=p_c$. The data exhibit a clear power-law regime over a broad range of cluster sizes, indicating scale invariance at criticality. Except $d=6$, which will be discussed in Sec.~\ref{sec-d6}, the size distributions are well described by
\begin{equation}
n(s,L) = s^{-\tau_{\mathrm{SCC}}} \tilde{n}(s/L^{d_{\mathrm{SCC}}}),  \label{eq-ns}
\end{equation}
with the Fisher exponent $\tau_{\mathrm{SCC}}$ consistent with the hyperscaling relation
\begin{equation}
\tau_{\mathrm{SCC}} = 1 + \frac{d}{d_{\mathrm{SCC}}} = 1 + \frac{1}{d^*_{\mathrm{SCC}}}.
\end{equation}

To further verify the scaling ansatz, we perform data collapse by plotting $s^{\tau_{\mathrm{SCC}}} n(s,L)$ as a function of the rescaled variable $s/L^{d_{\mathrm{SCC}}}$. As shown in the insets of Fig.~\ref{f8}, data for different system sizes collapse onto a single universal curve, confirming the validity of the scaling form of Eq.~\eqref{eq-ns}. These results demonstrate that SCCs exhibit conventional critical scaling behavior analogous to standard percolation, despite their higher-order connectivity constraints.

\subsubsection{Fractal dimensions of ICs and OCs}

We further find that both the giant IC and OC exhibit fractal geometry. Fitting the data to the finite-size scaling form in Eq.~(\ref{eq_fit}), their fractal dimensions are consistent with that of standard percolation clusters, i.e., $d_{\mathrm{IC}} = d_{\mathrm{OC}} = d_f$, as summarized in Table~\ref{taball}. This equality of the two fractal dimensions follows from the statistical symmetry between the giant IC and OC, as sketched in Fig.~\ref{f1}.

For finite systems, the probability that a site belongs to the giant IC or OC scales as
\begin{equation}
 \frac{C_1(\mathrm{IC)}}{V} = \frac{C_1(\mathrm{OC})}{V}\sim L^{d_f-d}.
\end{equation}
Since the intersection of the giant IC and OC corresponds to the giant SCC, a site belongs to the giant SCC if and only if it belongs to both the giant IC and OC. If the giant IC and OC are statistically independent, we then obtain
\begin{align}
\frac{C_1(\mathrm{SCC})}{V} &= \frac{C_1(\mathrm{IC)}}{V} \frac{C_1(\mathrm{OC})}{V}  \nonumber \\
&\sim L^{2d_f-2d},
\end{align}
which implies
\begin{equation}
d_{\mathrm{SCC}} = 2d_f - d.
\end{equation}
Table~\ref{taball} shows that this relation holds only for $d \geq d_u = 6$, indicating that IC and OC membership becomes effectively uncorrelated above the upper critical dimension. Below $d_u$, correlations between the two membership events enhance their overlap, resulting in denser SCCs and hence $d_{\mathrm{SCC}} > 2d_f - d$.

\subsection{Logarithmic corrections at $d_u=6$}  \label{sec-d6}

\begin{figure}
\centering
\includegraphics[width=\columnwidth]{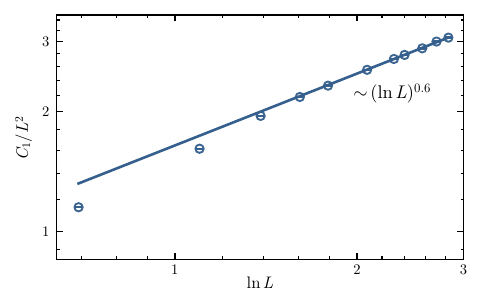}
\caption{The rescaled size of the largest SCC, $C_1/L^2$, as a function of $\ln L$ in dimension $d=6$. The line indicates the scaling $C_1/L^2\sim(\ln L)^w$ with $w\approx0.6$, suggesting multiplicative logarithmic corrections in the finite-size scaling of $C_1$.}    \label{f9}
\end{figure}

At the upper critical dimension $d_u=6$, the SCCs are expected to exhibit the mean-field behavior $d_{\mathrm{SCC}}/d=1/3$~\cite{Luczak1990,Luczak2009}, giving $d_{\mathrm{SCC}}=2$. This leading mean-field scaling is expected to be accompanied by multiplicative logarithmic corrections at $d=d_u$~\cite{Essam1978,Nakanishi1980,RuizLorenzo1998,Kenna2006}. To visualize the multiplicative logarithmic correction, we plot $C_1/L^2$ as a function of $\ln L$ in Fig.~\ref{f9}. The trend at large $L$ suggests a scaling form $C_1/L^2\sim(\ln L)^w$. Accordingly, the finite-size scaling ansatz in Eq.~\eqref{eq_fit} is modified to
\begin{equation}
C_1(\mathrm{X})=L^{d_\mathrm{X}}(\ln L)^w \left(a_0+a_1L^{-\omega_1}+\cdots\right), \label{eq_fitln}
\end{equation}
to incorporate the multiplicative logarithmic correction. Allowing all parameters to vary freely leads to unstable fits. Instead, fixing $a_1=0$ and $w=0.6$, as suggested by Fig.~\ref{f9}, we obtain $d_{\mathrm{SCC}}=1.998(8)$, as reported in Table~\ref{taball}.

With this multiplicative logarithmic correction, we define the rescaled largest SCC size in Fig.~\ref{f6}(e) as $x\equiv C_1/[L^2(\ln L)^{0.6}]$. The resulting distribution $P(x)$ exhibits an improved data collapse compared with the pure power-law rescaling $x\equiv C_1/L^2$, providing further evidence for the multiplicative logarithmic correction with $w\approx0.6$.

Furthermore, the size distribution at the upper critical dimension also contains a logarithmic correction~\cite{Essam1978,Nakanishi1980}. For SCCs, it is expected to take the form
\begin{equation}
n(s,L) =s^{-\tau_{\mathrm{SCC}}}(\ln s)^\theta \tilde{n}\left(\frac{s}{L^{d_{\mathrm{SCC}}}(\ln L)^w}\right). \label{eq-nsln}
\end{equation}
The factor $(\ln s)^\theta$ accounts for the deviation of the $d=6$ data from the pure power law $n(s,L)\sim s^{-4}$ observed in Fig.~\ref{f8}(e). In the inset of Fig.~\ref{f8}(e), we compensate for this logarithmic correction by plotting $n(s,L)s^{\tau_{\mathrm{SCC}}}(\ln s)^{-\theta}$ against $s/[L^{d_{\mathrm{SCC}}}(\ln L)^w]$. A good data collapse is obtained for different system sizes, with $\theta\approx1$.

The logarithmic corrections observed in $C_1$, the largest-SCC size distribution $P(x)$, and the SCC-size distribution $n(s,L)$, shown in Figs.~\ref{f9}, \ref{f6}(e), and \ref{f8}(e), provide strong support for $d_u=6$. For standard percolation, the corresponding exponents are $w=4/21\approx0.19$~\cite{RuizLorenzo1998} and $\theta=2/7\approx0.28$~\cite{Essam1978,Nakanishi1980}. The substantially different values observed here indicate that SCCs exhibit distinct correction-to-scaling behavior at the upper critical dimension.

\subsection{Complete graphs}

\begin{figure}
\centering
\includegraphics[width=\columnwidth]{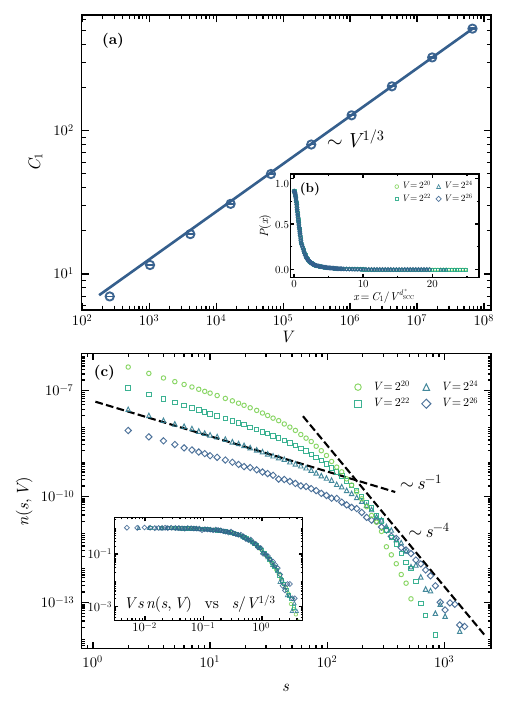}
\caption{Finite-size scaling of SCCs on complete graphs at criticality. (a) The size of the largest SCC, $C_1$, as a function of the system volume $V$. The solid line indicates the scaling $C_1 \sim V^{1/3}$. The inset shows the distribution $P(x)$ of the largest SCC. Data for different system sizes collapse when plotted as $P(x)$ versus $x\equiv C_1 / V^{d_{\mathrm{SCC}}^*}$, with $d_{\mathrm{SCC}}^* = 1/3$. (b) The size distribution $n(s,V)$ exhibits a double-scaling behavior. Large SCCs are governed by the Fisher exponent $\tau_{\mathrm{SCC}} = 1 + 1/d_{\mathrm{SCC}}^* = 4$, while small SCCs follow a distinct scaling $n(s,V) \sim V^{-1} s^{-\tau'}$ with $\tau' = 1$. The inset shows the corresponding scaling collapse using $V s^{\tau'} n(s,V)$ versus $s/V^{d_{\mathrm{SCC}}^*}$.}  \label{f10}
\end{figure}

We now turn to SCC percolation on complete graphs, which represents the mean-field limit. In this case, the absence of an intrinsic geometric length scale requires finite-size scaling to be formulated in terms of the system volume $V$.

Figure~\ref{f10}(a) shows the mean size of the largest SCC at criticality. The data are consistent with
\begin{equation}
C_1 \sim V^{d_{\mathrm{SCC}}^*},
\end{equation}
with $d_{\mathrm{SCC}}^*=0.3336(4)$, in agreement with the mean-field prediction $d_{\mathrm{SCC}}^*=1/3$~\cite{Luczak1990,Luczak2009}.

To further substantiate this scaling, we examine the size distribution of the largest SCC. As shown in the inset of Fig.~\ref{f10}(a), data for different system sizes collapse onto a single curve when plotted as $P(x)$ versus $x \equiv C_1 / V^{d_{\mathrm{SCC}}^*}$ with $d_{\mathrm{SCC}}^*=1/3$. This provides strong evidence that the largest SCC in infinite dimensions follows the standard finite-size scaling form, consistent with that observed in finite dimensions.

However, this does not imply that the behavior of SCCs on complete graphs is identical to that on hypercubic lattices above the upper critical dimension $d_u=6$. As we show in the next subsection, the finite-size behaviors of SCCs in $d>6$ do not fully coincide with those on complete graphs.

\subsection{Cluster statistics above $d_u=6$}

We now compare the finite-size behaviors of SCCs in $d=7$ with those on complete graphs. Although both systems lie above the upper critical dimension and share the same mean-field fractal dimension $d_{\mathrm{SCC}}^*=d_{\mathrm{SCC}}/d = 1/3$, their finite-size properties exhibit qualitative differences.

Figure~\ref{f8}(f) shows that in $d=7$, the size distribution $n(s,L)$ follows the same scaling form as in lower dimensions, characterized by a single power-law regime $n(s,L) \sim s^{-\tau_{\mathrm{SCC}}}$, with a cutoff at $s \sim L^{d_{\mathrm{SCC}}}$. The corresponding data collapse shown in the inset confirms that the standard finite-size scaling picture remains valid.

In contrast, the behavior on complete graphs is qualitatively different. As shown in Fig.~\ref{f10}(b), the distribution $n(s,V)$ exhibits a double-scaling structure. While the distribution of the large SCCs follows the mean-field scaling with $\tau_{\mathrm{SCC}}= 1 + 1/d_{\mathrm{SCC}}^* = 4$, small SCCs obey a distinct scaling
\begin{equation}
n(s,V) \sim V^{-1} s^{-\tau'}, \qquad \tau' = 1.  \label{eq_nscg}
\end{equation}
Here the factor $V^{-1}$ means that this contribution vanishes as $V\to\infty$, which is confirmed by the inset of Fig.~\ref{f10}(b).

\begin{figure}
\centering
\includegraphics[width=\columnwidth]{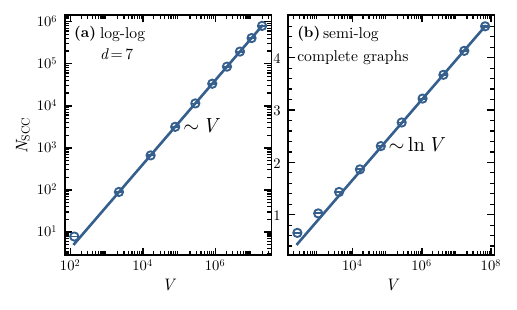}
\caption{Mean number of SCCs, $N_{\mathrm{SCC}}$, for (a) $d=7$ and (b) complete graphs. (a) For $d=7$, the data follow $N_{\mathrm{SCC}} \sim V$, indicating that $n(s,L)$ approaches a finite limit for fixed $s$ as $L \to \infty$. (b) For complete graphs, the number of SCCs scales as $N_{\mathrm{SCC}} \sim \ln V$.}    \label{f11}
\end{figure}

This difference is further reflected in the total number of SCCs. As shown in Fig.~\ref{f11}, one finds $N_{\mathrm{SCC}} \sim V$ in $d=7$, whereas $N_{\mathrm{SCC}} \sim \ln V$ on complete graphs. This logarithmic growth for complete graphs and the distribution Eq.~\eqref{eq_nscg} can be understood consistently from the scaling form of $n(s,V)$.

Based on Eq.~\eqref{eq_nscg} and the observation in Fig.~\ref{f10}(b), we assume a scaling form of $n(s,V)$ for complete graphs,
\begin{equation}
n(s,V) = V^{-h} s^{-\tau'} \tilde{n}(s / V^{d_{\mathrm{SCC}}^*}),
\end{equation}
where $V^{-h}$ is for the vanishing of $n(s,V)$ as $V\to\infty$, and $\tau'$ is the Fisher exponent for small SCCs. According to the definition of cluster-size distribution, one obtains the total number of SCCs,
\begin{align}
N_{\mathrm{SCC}} &= V \int n(s,V)\, ds   \nonumber \\
&\sim V^{1-h} \int^{V^{d_{\mathrm{SCC}}^*}}_{s=1} s^{-\tau'} ds.
\end{align}
To recover the scaling $N_{\mathrm{SCC}} \sim \ln V$, it requires $\tau'=1$ and $h=1$, consistent with the numerical data in Fig.~\ref{f10}(b).

This difference between hypercubic lattices and complete graphs can be understood from their underlying geometrical structures. At criticality, complete graphs ($p_c \sim 1/V$) are locally tree-like: connected clusters are dominated by tree structures, with only a finite number of loops, and the total number of loops in the system scales as $\sim \ln V$~\cite{Bollobas2001}. In contrast, hypercubic lattices retain an underlying spatial structure, where loops exist on multiple length scales due to the presence of short-range connections. As a consequence, the number of SCCs in finite-dimensional systems scales extensively with the system volume, $N_{\mathrm{SCC}} \sim V$. We point out that similar finite-size behaviors have also been observed in other systems~\cite{BenNaim2005,Hu2016,Huang2018,Yang2024}.

\section{Discussion}   \label{sec:discussion}

In this work, we have investigated the percolation transition of SCCs in systems with directed bonds. Using large-scale simulations in dimensions $d=2$-$7$ and on complete graphs, we determined the fractal properties of SCCs and related structures. For $d < d_u=6$, SCCs exhibit nontrivial fractal dimensions, with $d_{\mathrm{B}} < d_{\mathrm{SCC}} < d_f$, indicating that SCCs form a distinct class of connectivity structures, intermediate between the backbone and the full percolation cluster. Furthermore, unlike $d_f$, which increase monotonically with $d$, $d_{\mathrm{SCC}}$ peaks at $d=4$, indicating a different dimensional dependence. Besides, the giant IC and OC are also found to be fractal with dimensions consistent with standard percolation.

Although the fractal dimension takes mean-field value for $d \geq d_u=6$, the full cluster statistics retain signatures that distinguish finite-dimensional systems from the infinite-dimensional limit. In $d=7$, SCCs follow conventional finite-size scaling, characterized by a power-law cluster-size distribution, $\sim s^{-4}$, and an extensive number of clusters, $N_{\mathrm{SCC}} \sim V$. In contrast, on complete graphs, SCCs exhibit different finite-size behaviors: the number of SCCs grows only logarithmically, $N_{\mathrm{SCC}} \sim \ln V$, and the cluster-size distribution displays a double-scaling form, with $\sim s^{-1}$ for small SCCs and $\sim s^{-4}$ for large SCCs.

Our work provides a unified picture of SCC percolation across dimensions and highlights several features that call for further investigation. In particular, $d=4$ appears to be a special dimension: $d_{\mathrm{SCC}}$ reaches its maximum there, while the susceptibility-like quantity $\chi$ ceases to diverge above $d=4$. A theoretical understanding of these features and their relation to the crossover toward mean-field behavior at $d_u=6$ remains an open problem. Another open question concerns directional asymmetry. The present results rely on the statistical symmetry between incoming and outgoing connectivity. It would be interesting to investigate how breaking this symmetry changes the critical behavior and whether it establishes a connection to directed percolation, whose upper critical dimension $d_u=4$ differs from the $d_u=6$ found here.

\section*{Acknowledgements}

The research was supported by the National Natural Science Foundation of China under Grant No.~12675037.

\bibliography{sccRefs}

\end{document}